\documentclass[prd,aps,preprintnumbers,showpacs,10pt]{revtex4}
\usepackage{amsfonts,amsmath,amssymb,bm,dsfont,eucal}
\usepackage[english]{babel}
\usepackage[latin1]{inputenc}
\usepackage[T1]{fontenc}
\usepackage{ae,aecompl}
\usepackage{graphicx,graphics,epsfig,epstopdf,subfigure,color,psfrag}
\usepackage{hyperref}

\def \a {\alpha}

\def \d {\delta}

\def \m {\mu}
\def \n {\nu}

\def \non {\nonumber}

\def\laq{~\raise 0.4ex\hbox{$<$}\kern -0.8em\lower 0.62ex\hbox{$\sim$}~}
\def\gaq{~\raise 0.4ex\hbox{$>$}\kern -0.7em\lower 0.62ex\hbox{$\sim$}~}
\newcommand{\be}{\begin{equation}}
\newcommand{\ee}{\end{equation}}
\newcommand{\bea}{\begin{eqnarray}}
\newcommand{\eea}{\end{eqnarray}}
\newcommand{\nn}{\nonumber}

\begin{document}
\preprint{BARI-TH/612-2009}
\preprint{DCPT/09/128}
\preprint{IPPP/09/64}
\title{ Holographic Approach to Finite Temperature QCD: \\ The Case of  Scalar Glueballs and  Scalar Mesons}
\author{Pietro Colangelo}  
\email{pietro.colangelo@ba.infn.it}
\affiliation{Istituto Nazionale di Fisica Nucleare, Sezione di Bari, Italy}
\author{Floriana Giannuzzi}
\email{floriana.giannuzzi@ba.infn.it}
\affiliation{Dipartimento di Fisica dell'Universit\'a di Bari and INFN, Sezione di Bari, Italy}
\author{Stefano Nicotri}
\email{stefano.nicotri@durham.ac.uk}
\affiliation{Institute for Particle Physics Phenomenology, Department of Physics, Durham University,
Durham DH1 3LE, United Kingdom}
\begin{abstract}
We study   scalar glueballs and scalar mesons at $T\neq 0$ in the soft wall holographic QCD model. We find that, using the Anti-de Sitter-Black Hole metric for all values of the temperature,
the masses of the hadronic states decrease and the widths become broader when $T$ increases, and there are  temperatures for which the states disappear from the scalar glueball and scalar meson spectral functions. 
However, the values of the temperatures in correspondence of which such phenomena occur are  low, of the order of $40-60$ MeV. A  consistent holographic description of  in-medium  effects 
on hadron properties should include the Hawking-Page transition, which separates the phase with  the Anti-de Sitter metric at small temperatures from the phase with Anti-de Sitter-Black Hole metric at high temperatures.
\end{abstract}

 \maketitle

\section{Introduction}
Finite temperature effects on hadron properties are currently investigated in heavy ion collision experiments, such as the ones at the Relativistic Heavy Ion Collider (RHIC) at Brookhaven, and will be the subject of analyses 
at the CERN Large Hadron Collider \cite{rassegna}.  It is presently believed that the hot and dense  medium where the hadrons are created modifies masses and  widths, and 
that this distortion can be observed reconstructing the states  produced in the hadron decays.  A significant example is represented by heavy mesons like  $J/\psi$ and the radial and orbital $c \bar c$ excitations:
it is expected that these mesons have different behaviour when the temperature of the medium  varies, with the lowest lying state ($J/\psi$)  surviving in the medium after the deconfinement transition,
up to temperatures $T \simeq 1.5 \,T_c$,  while the excited states  melt  at temperatures close to the deconfinement temperature $T_c$ \cite{Ding:2009se}. 

The  theoretical tools used  to describe these phenomena  are lattice QCD, QCD Sum Rules, effective QCD theories and models of QCD \cite{rassegna1}.  Recently,
 in the framework of the  gauge/string duality approach \cite{Maldacena:1997re,Witten:1998qj,Gubser:1998},
 it has been suggested  that a description of a strongly coupled quantum gauge theory at finite temperature can be obtained from a semiclassical theory formulated into a higher dimensional Anti-de~Sitter (AdS) spacetime containing a black hole (BH)  \cite{Witten:1998zw}.  In order to achieve smoothness and completeness of the metric, also the time direction must be compact \cite{Witten:1998zw}; the temperature is inversely proportional to the distance $z_h$ of the black hole horizon to the AdS boundary: $T=1/(\pi z_h)$. Therefore, the smallest is the system, the highest is the effect of the thermal fluctuations.  

Another way to describe finite temperature effects in a  holographic description is to consider a thermal AdS space, i.e. an Anti-de~Sitter space with a compact Wick-rotated time, with no black hole. In this case, as in standard finite temperature field theory, the temperature is given by the inverse dimension of the compactified time direction.

In  Ref.\cite{Witten:1998zw} the gravity dual of ${\CMcal N}=4$~SYM theory on $S^3\times S^1$, with  $N \to \infty$,  has been studied, showing the existence of two phases. A critical temperature was found  at which a first order Hawking-Page (HP) transition \cite{HPage} occurs between the thermal AdS and the Anti-de Sitter-Black Hole (AdS-BH)  geometries, due to an inversion of the hierarchy of the free energies relative to the two metrics \footnote{A finite temperature phase-transition can occur since conformal symmetry is broken by formulating the dual field theory  on a  spatial three-sphere, and the large $N$ limit is considered.}. The  temperature at which this transition occurs has been identified with  the  deconfinement  temperature of the gauge theory on the boundary
\footnote{For the gauge group $SU(N)$ and  large $N$ the deconfinement transition is found to be of first order \cite{teper}.},  which in this way receives a remarkable geometric interpretation.

A  description inspired to gauge/string duality has been applied to  QCD  considering, in particular,  two holographic models constructed in the  (phenomenological) so-called bottom-up approach
\footnote{For a holographic description of  finite temperature QCD in the top-down approach see \cite{erdmenger}
and references therein.}:  the hard wall (HW) and the soft wall (SW) model.  In these two models, QCD is the gauge theory defined on the boundary of a five-dimensional holographic space. Since  gauge/gravity 
(Anti-de Sitter/Conformal Field Theory - AdS/CFT) correspondence
has been conjectured for a conformal theory defined on the boundary, the generalization  of duality to QCD (which is not a conformal theory) requires a mechanism to break conformal invariance. The HW and SW models
differ in the way  this invariance is broken: a sharp cut-off of the holographic space, up to a maximum value of the fifth coordinate $z_m$, is imposed  in the HW model, with  $z_m={\cal O}(\Lambda^{-1}_{QCD})$  \cite{polchinsky,son1,pomarol1,teramond1,Grigoryan:2007vg};   a background dilaton-like field, vanishing at the AdS boundary and involving a dimensionful scale $c={\cal O}(\Lambda_{QCD})$ , is included  in the soft wall model, implementing a smooth cut-off of the  holographic space
\cite{Karch:2006pv,Andreev:2006vy}.  At $T=0$ the two models allow to compute  QCD quantities such as the spectrum and decay constants of light mesons and glueballs, meson form factors and strong couplings, QCD condensates:  in many cases the 
agreement with experiment is  noticeable. In  those models,   the metric is not a dynamical field, hence, when 
generalized to the case of finite temperature,  there is no  reason to justify the  occurence of a Hawking-Page phase transition of the kind described in \cite{Witten:1998zw}.  However, the existence of such  transition has been  inferred \cite{Herzog:2006ra,braga}: considering the free energies of the thermal AdS and AdS-BH configurations,   in both the models it was found that the stable metric is thermal AdS at low temperatures and AdS-BH at high temperatures.  The
 temperature at which the transition occurs depends on the model:    $\displaystyle T_{HP}={2 ^{1/4} \over \pi z_m}$ in the case of the hard wall model,  $z_m$ being  the position of the hard wall;  
$\displaystyle T_{HP}={1 \over \pi } {c \over 0.647}$ in the case of the soft wall model, $c$ being the dimensionful constant introduced by the background dilaton field in the model. Using  the value for $c$ determined at zero temperature,
in the SW model the HP  temperature  is $\displaystyle T_{HP}={1 \over 2 \pi } {m_\rho \over  0.647}\simeq 192$ MeV,    $m_\rho$  being the mass of the $\rho$ meson \cite{Herzog:2006ra}.  Therefore, the Hawking-Page transition temperature $T_{HP}$  is  close to the QCD deconfinement temperature obtained, e.g., by lattice QCD simulations. The presence of the HP transition  has been found also in different holographic models of QCD  \cite{Gursoy:2009jd}.

An analysis of vector mesons at  $T\neq 0$ has been carried out  in the SW model  with AdS-BH metric for all values of $T$, and the temperature dependence of vector meson masses and widths has been investigated \cite{Fujita:2009wc}. In this analysis,  the  scale fixing the physical  temperature, $c_{J/\psi}$, which  appears once again in the dilaton field,  has been suitably chosen   to  obtain   the masses  of hidden charm vector  mesons
($J/\psi$ and radial excitations) at $T=0$.  Thus,  it is assumed that the model describes $c \bar c$   in a thermalized medium, and   a critical temperature has been identified,  where the lightest  charmed  vector  meson disappears from the two-point spectral function of  the retarded two-point Green's function of  the vector operator $J_\mu=\bar c \gamma_\mu c$ in the boundary theory  \cite{Fujita:2009wc}.

Here we  consider an analogous problem: we investigate the SW model  with an AdS-BH  metric for all values of $T$,  thus including the effects of the temperature with no reference to the HP transition, focusing on the cases of scalar glueballs and  scalar mesons.  We show that, in both cases,   the qualitative dependence on the temperature of the hadron masses and widths  follows general expectations, the masses becoming lighter and the widths broader than at $T=0$.  However, the physical values of 
the temperature  at which  such phenomena occur are  lower  than  found  in   determinations based, e.g.,  on lattice QCD simulations,  and occur in the confined phase of QCD, so that the use of  AdS-BH metric for all values of temperature    is  inappropriate.
In order to describe QCD at finite $T$  by this model,  the  Hawking-Page transition must be taken into account,  using  the thermal AdS metric for $T<T_{HP}$ and the  AdS-BH metric for $T>T_{HP}$:
we consider this case,  finding  a simple behaviour of the hadron masses and widths  versus $T$,  different from the behaviour found using  AdS-BH  for all  $T$. 

The plan of the paper is the following. In Sect. \ref{sect:glueball} we consider  the sector of scalar glueballs at $T \neq 0$, with   the AdS-BH metric in the soft wall model. We discuss the temperature dependence of the mass and width, 
finding  the values of the temperatures where the states dissolve in the equilibrated medium.  
In Sect.\ref{sect:scalar}  we find the same features in the sector of  light  scalar mesons described in the SW AdS-BH model, finding that the scalar meson spectral function does not reveal  resonances  already at low temperatures.  In Sect.\ref{sec:Hawking-Page} we discuss  the model  including the Hawking-Page transition:
now the spectral functions do not display peaks at $T\geq T_{HP}$, therefore hadronic states do not survive the deconfinement transition.
In the appendix we discuss the scalar glueball in the hard wall model at $T\neq0$.  

\section{Scalar glueball at $T \neq 0$:   soft wall  model with AdS-BH metric}\label{sect:glueball}
We start our discussion ignoring the issue of the stability of the metric in different ranges of the   temperature, and consider the case  of a dual QCD model described, for all values of  $T$, by 
an Anti-de Sitter-Black Hole  metric,  defined by the metric tensor:
\begin{equation}
g_{MN}=\left({ R \over z }\right)^2 diag\,\left(f(z),-1,-1,-1,-1/f(z)\right)\,\,\, 
\end{equation}
($M,N=0,1,2,3,5$),  where $R$ is the AdS radius and  
\begin{equation}
 f(z)=1- \left({z \over z_h}\right)^4 \,\,\, .
\end{equation}
$z_h$ is the position of the  black-hole horizon along the holographic axis  $z$,  and is related to the Hawking temperature $T$:  $T=1/(\pi z_h)$. The fifth coordinate $z$ varies in the range $0<z<z_h$.
In the $5d$ space  we define a field $X(x,z)$ dual to the QCD operator  $\CMcal{O}_{G}=\beta(\a_s)\,G_{\m\n}^a\,G^{\m\n\,a}$ ($a$ is color index) with   $J^{PC}=0^{++}$ and conformal dimension
$\Delta=4$; $\beta(\a_s)$ is the Callan-Symanzik function. This operator is defined in the field theory living on the $4d$
boundary. According to the AdS/CFT correspondence, the conformal dimension of a ($p$-form) operator on the boundary is related to
the AdS mass $m_5$ of the dual field in the bulk by the relation \cite{Witten:1998qj}:
\begin{equation}
m_5^2 R^2=(\Delta-p)(\Delta+p-4)\;.\label{m5}
\end{equation}
Following  the AdS/QCD correspondence dictionary, the  dual $X(x,z)$ of the scalar ($p=0$) field $\CMcal{O}_{G}$ is massless: $m_5^2=0$ 
\cite{Colangelo:2007pt}. It can be described, 
in the soft wall model,   by the $5d$ action $S_G$:
\begin{equation}\label{action}
S_{G}=\frac{1}{2k} \int d^4x \,dz  \,\, e^{-\phi(z)} \sqrt{g} \, g^{MN} \left( \partial_M X(x,z) \right) \left(\partial_N X(x,z)\right)
\end{equation}
where $k$ is a parameter  rendering  $S_G$ dimensionless,  and $g$ is the determinant of the metric tensor.  

The field  $\phi$ in eq.(\ref{action})  characterizes the holographic  model: it represents a background dilaton field depending on the coordinate $z$ only, and vanishing at the AdS boundary $z=0$. 
It allows to introduce a dimensionful
parameter $c$ producing a  breaking of conformal invariance.  In the model proposed in  \cite{Karch:2006pv} $\phi$ is given by the expression:
\begin{equation}\label{dilaton}
\phi(z)= (c  z)^2
\end{equation}  
and allows to obtain, at $T=0$, linear Regge trajectories for vector and axial-vector mesons \cite{Karch:2006pv}, scalar glueballs \cite{Colangelo:2007pt} and light scalar mesons \cite{Colangelo:2008us}.
The parameter $c$ fixes the hadronic scale: at $T=0$ it can be determined from the spectrum of the vector mesons, in particular from the $\rho$ mass, and is given by \cite{Karch:2006pv}:
\begin{equation}\label{crho}
c \equiv c_\rho= {m_\rho \over 2}=390  \,\,\, {\rm MeV} \,\,\, ,
\end{equation}
since the relation found for the masses of the vector mesons is: $m^2=4(n+1) c^2$, with $n=0,1,\dots$\,\,\, .
We keep the value \eqref{crho} in the discussion below.

In order to determine the $T$ dependence of the mass spectrum associated to the operator $\CMcal{O}_{G}$, we follow the  method used in  \cite{Fujita:2009wc}, and consider
the  equation of motion for the field $X$ obtained from  (\ref{action}):
\begin{equation}\label{eom}
e^{-\phi(z)} \sqrt{g}\, g^{\mu \nu} \partial_\mu \partial_\nu  X(x,z)+
\partial_z \left(  e^{-\phi(z)} \sqrt{g}\, g^{zz} \partial_z X(x,z) \right) =0 
\end{equation}
($\m,\n=0,1,2,3$). 
According to the AdS/CFT dictionary, 
field/operator duality implies that  the function $X_0(x)$ is associated to  the field  $X(x,z)$ in the AdS space, such that $X_0(x)$   acts, in the generating functional of the boundary theory, as the source of the  four dimensional (gauge invariant) local operator $O_G(x)$ dual to $X$. 
This implies the definition of the bulk-to-boundary propagator $K$ allowing to relate the field $X(x,z)$ on the bulk to its value on the boundary $X_0(x^\prime)$:
\begin{equation}\label{bulk-to-boundary}
X(x,z)=\int d^4x^\prime K(x,z;x^\prime, 0) X_0(x^\prime) \,\,\,\, ,
\end{equation}
with the condition  $K(x,z;x^\prime, 0) \xrightarrow[z\rightarrow0] {} \delta^4(x-x^\prime)$.
In the  momentum space the relation involving the bulk-to-boundary propagator $\tilde K$, the field $\tilde X$ and the source $\tilde X_0$ is:
\begin{equation}\label{btbp}
\tilde{X}(q,z)=\tilde K(q,z) \tilde X_0(q)
\end{equation}
with  $\tilde K(q,0)=1$.

From Eq.~\eqref{btbp} and Eq.~\eqref{eom},  one finds that $\tilde K(q,z)$ satisfies the equation:
\begin{equation}\label{eom2}
\tilde K''(q,z)- \frac{4-f(z)+2c^2z^2f(z)}{z f(z)} \tilde K'(q,z)+\left( \frac{q_0^2}{f(z)^2}-\frac{\bar{q}^2}{f(z)} \right)  \tilde K(q,z)=0 
\end{equation}
where $q=(q_0,\bar q)$ and  the primes denote derivatives with respect to the holographic variable $z$. This equation must be solved for different values of $q_0$ and $\vec q$, and we 
  first consider the case of  vanishing three-momentum $\vec q=0$.  Putting  $\omega=q_0$,  Eq. \eqref{eom2} can be written in terms of the dimensionless variable $u=z/z_h$:
\begin{equation}\label{eom3}
\tilde K''(\omega^2,u)- \frac{3+u^4+2c^2z_h^2 u^2 (1-u^4)}{u (1-u^4)} \tilde K'(\omega^2,u)+ \frac{\omega^2  z_h^2}{(1-u^4)^2}  \tilde K(\omega^2,u)=0 
\end{equation}
with the primes now denoting derivatives with respect to  $u$.

Let us specify the boundary conditions for (\ref{eom3}). The solution for  $u\to 0$  reads:
\begin{equation}\label{asy0}
\tilde K(\omega^2,u) \xrightarrow[u\rightarrow0]{} A(\omega^2) \left(1+\frac{\omega^2  z_h^2}{4} u^2+...\right)+ B(\omega^2) \left(\frac{c^4 z_h^4}{2} u^4+...\right)
\end{equation}
so that the condition $\tilde K(\omega^2,0)=1$  fixes   the coefficient  $A$: $A(\omega^2)=1$. For the second boundary condition we look at the solutions  of (\ref{eom3})
 near the horizon $u=1$: 
\begin{equation}\label{asy1}
\tilde K_\mp(\omega^2,u)=(1-u)^{\mp {i \sqrt{\omega^2 z_h^2}\over 4}} \,.
\end{equation}
As discussed in  \cite{starinets},  the choice of the boundary condition at the horizon selects the Green function obtained using the AdS/ CFT procedure in the Minkowskian space-time. 
We impose as a boundary condition the matching of $\tilde K$ with the {\it in falling } solution   of (\ref{eom3})  near the black-hole horizon:
\begin{equation}\label{fallingin}
\tilde K \xrightarrow[u\rightarrow1] {}\tilde  K_-(\omega^2,u) \sim(1-u)^{- {i \sqrt{\omega^2 z_h^2}\over 4}} \,,
\end{equation}
so that the matching of the boundary conditions of $\tilde K$ for $u\to 0$ and $u \to 1$ fixes the coefficient function $B(\omega^2)$, which can be determined numerically \cite{Teaney:2006nc}.

To get the glueball masses, we consider the relation allowing to compute,
within  the AdS/CFT correspondence,  correlation functions in the gauge theory defined on the boundary of the AdS space starting from  the effective action in the $5d$ bulk theory  
\cite{Witten:1998qj,Gubser:1998}  extended to the Minkowskian space-time:
\begin{equation}\label{generating}
  \biggl\langle e^{i\int d^4 x\;X_0(x )\,\CMcal{O}_{G}(x)}\biggr\rangle_{CFT}=
  e^{iS_{5d}[X(x,z)]} \,\,\, .
\end{equation}
In our case,  $S_{5d}[X(x,z)]$ is the  action
\eqref{action} of the  bulk field $X(x,z)$ dual to  $\CMcal{O}_{G}(x)$,  $X_0(x)$ being a source term.  As discussed in \cite{starinets} in case of Minkowskian correlators, the retarded two-point Green's function $\Pi^R_G$ can be derived
differentiating the r.h.s of eq.(\ref{generating}) with respect to the source,  and imposing the boundary conditions discussed above.
In terms of the  the bulk-to-boundary propagator $\tilde K$,  $\Pi^R_G$ reads:
\begin{equation}\label{2pfunction}
\Pi^R_G(\omega^2)= \left. \frac{1}{2k}    \frac{R^3 f(u)}{u^3 z_h^4} e^{-\phi(u)} \tilde K(\omega^2,u) \partial_u \tilde K(\omega^2,u) \right|_{u=0}  \,\,\, .
\end{equation}
%
\begin{figure}[t]
\begin{center}
\includegraphics[width=9cm]{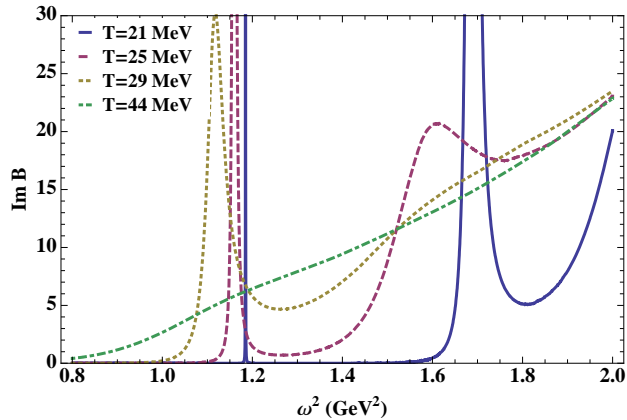}
\caption{Imaginary part of the coefficient $B(\omega^2)$, proportional to the spectral function Im$\Pi^R_G(\omega^2)$ of eq.(\ref{2pfunction}), at different temperatures $T$,  in the case of the scalar glueball and using  the  AdS-BH metric.}
\label{spfunction}
\end{center}
\end{figure}
Substituting eq.\eqref{asy0} in eq.\eqref{2pfunction}, we  determine the spectral function,  the imaginary part of $\Pi^R(\omega^2)$,  which is proportional to the imaginary part of the coefficient $B(\omega^2)$. 
The resulting  spectral function (modulo a numerical factor) is depicted in Fig. \ref{spfunction} for several values of the physical temperature obtained from the position $z_h$ of the BH horizon, using
 the value of $c$ in the dilaton background field fixed in \eqref{crho}.  
 
\begin{figure}[h]
\subfigure{
\includegraphics[width=6.5cm]{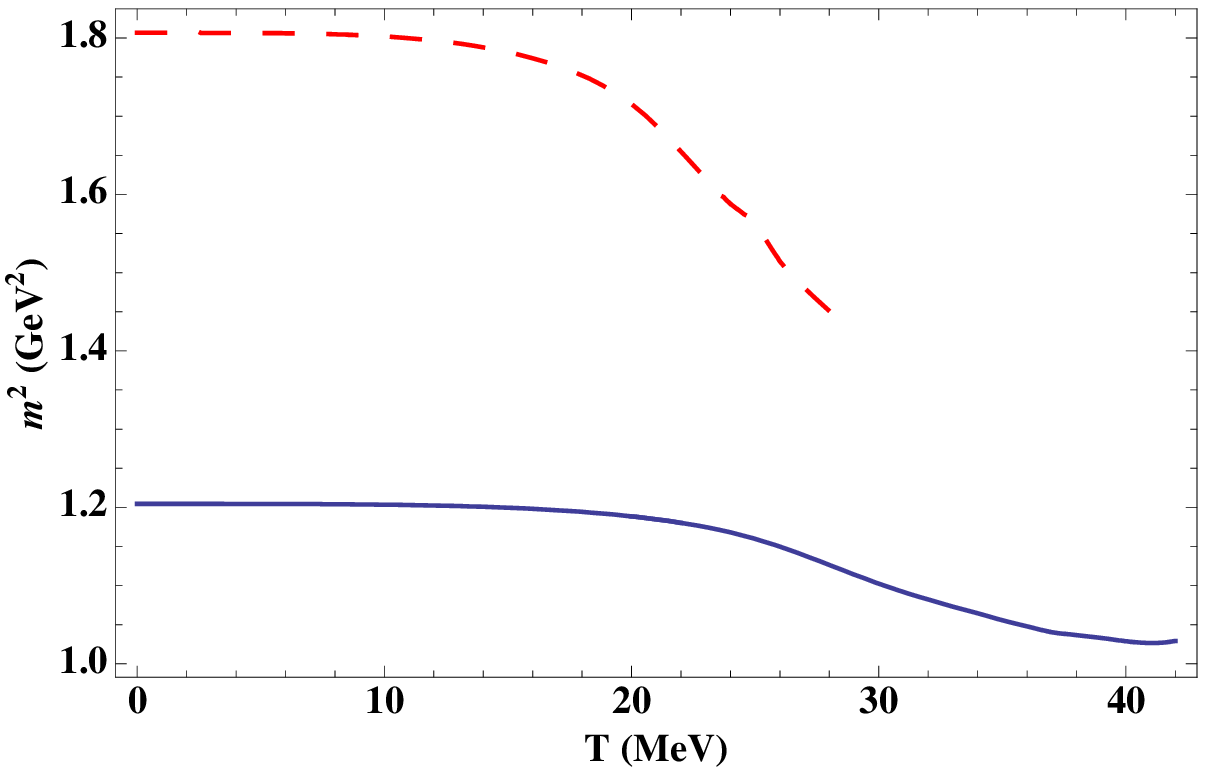}
\label{figmassT}
}
\hspace*{1cm}
\subfigure{
\includegraphics[width=6.5cm]{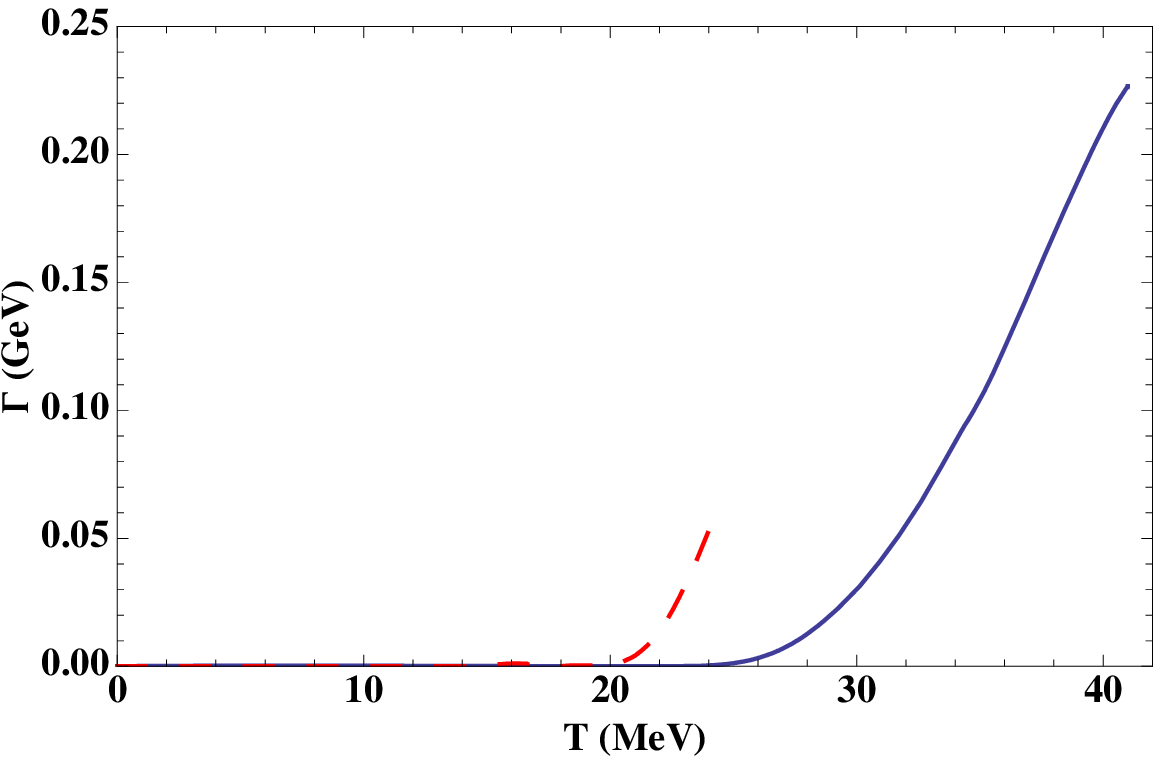}
\label{figwidthT}
}
\caption{Squared mass (left)  and width (right)  of the lightest  glueball state (continuous lines) and of the first   excited  state (dashed lines)  as a function of the temperature $T$ (MeV) in the SW model with  AdS-BH metric.
Each curve ends  at the temperature where the corresponding peak disappears  from the spectral function.}
\end{figure}

For each value of the temperature, the spectral function  in Fig. \ref{spfunction} displays various peaks, which become broader as $T$ increases. We identify the position of each peak with the mass of  scalar glueballs, in particular the lowest lying state and the first  excitation.  At  small values of  $T$:  $T<20$ MeV, the results  in \cite{Colangelo:2007pt} for the mass spectrum are recovered:  $m^2_G=(4 n+8)c^2$, i.e. $m^2_G=1.217$ GeV$^2$  and $m^2_G=1.825$ GeV$^2$ for the first two states. Increasing the temperature $T$ the position of the peaks is shifted towards smaller values and the widths become broader. Both these quantities can be determined fitting the spectral function
with a Breit-Wigner form \cite{Fujita:2009wc}:
\begin{equation}\label{breitwigner}
\rho(\omega^2)=\frac{a \, m \, \Gamma \, \omega^b}{(\omega^2-m^2)^2+m^2 \Gamma^2} 
\end{equation}
with parameters $a$ and $b$.
The results of $m^2$ and $\Gamma$  from the fit  are shown in Fig. \ref{figmassT} for the ground state and for the first excited state. At temperatures below  $T\sim20$ MeV ($T\sim17$ MeV for the excited state) the horizon of the black hole is far enough and the eigenfunctions vanish before reaching it, so that  in this range of temperatures it is possible to determine glueball masses as the eigenvalues of the  equation
\begin{equation}\label{eombog}
-H''(m^2,u)+\left( \frac{15}{4 u^2}+2c^2 z_h^2+c^4 z_h^4 u^2 \right)H(m^2,u)=m^2 z_h^2 H(m^2,u)
\end{equation}
coming from a Bogoliubov transformation of  Eq.(\ref{eom3})  as  at  $T=0$ \cite{Colangelo:2007pt};
in the range $T=20-22$ MeV  the results obtained solving the eigenvalue problem and fitting the spectral function coincide.

\begin{figure}[h]
\begin{center}
\includegraphics[width=9cm]{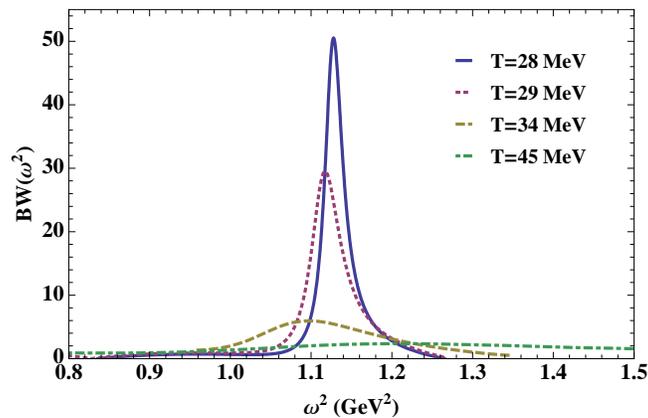}
\caption{Lowest-lying resonance  in Fig.\ref{spfunction}, after  subtracting from the spectral function  a term representing a continuum;  several temperatures are considered.}
\label{BW}
\end{center}
\end{figure}

At $T=25-30$ MeV  the squared glueball mass  is reduced to about 80$\%$ of its value at $T=0$:  the second peak disappears at $T\sim29$ MeV, while  the peak corresponding to the lowest lying state persists until  $T\sim45$ MeV. 
Let us discuss in more detail the structure corresponding to  the lightest scalar glueball.
At $T=28$ MeV the width of the first peak in the spectral function becomes sizeable, i.e. larger than $1\%$ of the corresponding mass.  We  fit  the spectral function by a
a  Breit-Wigner term plus a  function,  which we interpret as representing a continuum, of the simple form $P(x)=a+b\, x+c\, x^d$, with $1<d<2$ and $x=\omega^2$.
We define a melting temperature as the one at which the height of the Breit-Wigner peak obtained after subtracting the continuum contribution
from the spectral function is less than $0.05$ times the value at $T=28$ MeV; in this way we find temperature:  $T\approx 45$ MeV.  The result of subtracting the continuum is depicted in Fig. \ref{BW},
which shows the broadening and the shift of the mass of the resonance.  Near the melting temperature we also observe
a slight rising of the mass, analogously to what found in ref. \cite{Fujita:2009wc}. Using the same criterion, the melting temperature of the first excited state is $T\approx 29$ MeV.

The behaviour of the  width of the first two peaks  with respect to $T$ is shown in Fig. \ref{figwidthT}. It is  qualitatively 
analogous to the behaviour of the scalar glueball mass and  width observed in  lattice studies \cite{lattice}: however,  the temperature scale is very different. 

\begin{figure}[b]
\begin{center}
\includegraphics[width=7cm]{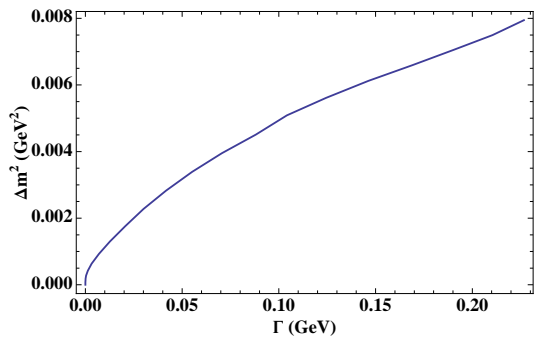}
\caption{Difference $\Delta m^2(T)=(m(T=0)-m(T))^2$ (GeV$^2$)   versus the width $\Gamma(T)$ (GeV)  varying the temperature $T$ in the range  $T=20-45$ MeV,  in the case of the lightest scalar  glueball.  }
\label{figDM2vsL}
\end{center}
\end{figure}

In \cite{Fujita:2009wc},  in the vector meson channel, a range of temperatures was found where   $\Delta m^2$ (with $\Delta m= m(T=0)-m(T)$) is linearly related to  the width $\Gamma(T)$:  in the case of the lowest lying scalar glueball, we  find an approximately linear dependence of $\Delta m^2$ on $\Gamma(T)$ in the range $T=20-45$~MeV, as depicted Fig. \ref{figDM2vsL}:  such a relation could represent a benchmark for other approaches to  finite T QCD.

\begin{figure}[h]
\begin{center}
\includegraphics[width=8cm]{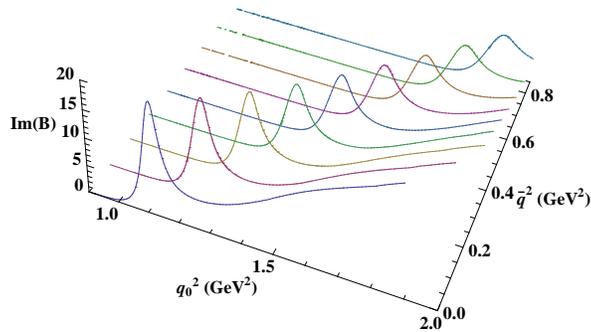}
\caption{Imaginary part of the coefficient $B(q_0^2,\vec q^2)$, proportional to the spectral function Im$\Pi^R_G(q_0^2,\vec q^2)$, for the scalar glueball at $T=30$ MeV, for different values of the three-momentum squared $\bar q^2$ in the range $\bar q\,^2=0 - 0.8$ GeV$^2$, in the SW model with  AdS-BH metric.}
\label{figimBqbar}
\end{center}
\end{figure}

For non-vanishing values of the three-momentum,  $\bar q\neq 0$,  the results are similar.
The spectral function can be obtained from eq. \eqref{eom2}, imposing the same boundary conditions as for  $\vec q =0$.
As depicted in Fig.~\ref{figimBqbar} at $T=30$ MeV and for  values of $\bar q^2$ in the range $\bar q^2=0-0.8$ GeV$^2$, increasing $\bar q^2$  the peaks of the spectral function are shifted towards higher values of $q_0^2$ and become broader. The difference $q_0^2-\bar q^2$ is not constant, Lorentz invariance being violated in the finite temperature theory.  The same effect,  shown in Fig.~\ref{figimBqbar},  was found in the case of vector mesons  \cite{Fujita:2009wc}. 

The qualitative behaviour of  the mass and width of scalar glueballs  in the soft wall model with AdS-BH metric is similar to the one found in \cite{Fujita:2009wc} for vector mesons:
increasing $T$ continuously from $T=0$, the mass of the various states is shifted to lower values while the width increases, and at some critical value of $T$ the peaks disappear from the spectral function.
The broadening of the states and their disappearance can inspire the picture of their melting in the thermalized medium; moreover, the temperature at which, e.g., the first excitation disappears from the spectral function (dissolves) is lower than the 
temperature at which the lowest lying state disappears, thus suggesting  that  lightest resonances persist in the medium at  higher  temperatures also in the case of glueballs. 

However, the physical temperatures at which such phenomena occur are  low ($30-40$~MeV),  the scale being determined by the dimensionful constant $c$ in the background dilaton field (\ref{dilaton}) and  fixed at $T=0$ to \eqref{crho}. Such low temperatures  are  inconsistent with those found, e.g., in lattice calculations of the glueball masses, in which a decrease of about $20\%$ is observed for the mass of the 
lightest scalar glueball in the range of temperatures $0.81\,T_c<T<T_c$,  with  $T_c\simeq 260$~MeV  \cite{lattice}.
In the case of vector mesons, in  \cite{Fujita:2009wc} a further  assumption was adopted:  the scale fixing the temperature in this  sector, called  $c_{J/\psi}$, is different from the scale $c=c_\rho$ fixed from the
$\rho$ vector meson spectrum, considering  mesons  of $c \bar c$ type  in the holographic approach. The resulting melting temperature of vectors can be compared to the deconfinement temperature in QCD:
since the ratio of the masses of $\rho$ and ${J/\psi}$  is recovered for  $c_{J/\psi}\simeq 4 c_\rho$, a value $c_{J/\psi}\simeq 1.56$ GeV produces values of the $J/\psi$  dissociation temperature  of about  $230$ MeV  \cite{Fujita:2009wc}.

In the case of glueballs, however, it is unclear how an analogous assumption   could be adopted:  there are no other scales  in terms of which the physical temperature can be expressed, 
  the only possibility being  $c=c_\rho$;  hence,  the low physical dissociation temperatures inferred from Figs. \ref{spfunction}-\ref{figimBqbar} is not avoidable. 

The problem  finds a solution if one considers  the stability of the metric, i.e.  the fact that at low temperature the stable metric is thermal AdS, not AdS-BH. However, before considering this issue, let us analyze
another sector of QCD, the light   scalar mesons.

\section{Scalar mesons at $T \neq 0$:   soft wall model with AdS-BH metric} \label{sect:scalar} 

The analysis of the scalar meson sector at $T\neq 0$ in the AdS-BH soft wall model produces  results analogous to the case of scalar glueballs.
We generalize the $5d$ action studied in  \cite{Karch:2006pv} and also considered in \cite{Fujita:2009wc}:
\begin{equation}
S_{eff}=\frac{1}{k^\prime}\int
d^4x dz \,e^{-\phi(z)}\, \sqrt{g}\,\, \mbox{Tr}\Big\{|D Y |^2-m_5^2 Y^2-\frac{1}{4g_5^2}\big(F_L^2+F_R^2\big)\Big\}
\label{action-new}
\end{equation}
which includes  fields  dual to  QCD operators defined at the boundary $z=0$.
There is a scalar bulk field  $Y$ of mass $m_5^2$,   written as
\begin{eqnarray}
Y=(Y_0+S)e^{2i\pi}
\end{eqnarray}
in terms of a background field $Y_0(z)$, of the  scalar
field $S(x,z)$ and of the chiral field $\pi(x,z)$.  $Y_0$ is dual to $\langle \bar q q\rangle$ ($q$ are light quarks) and  represents the term responsible  of the chiral symmetry breaking \cite{son1,pomarol1,Karch:2006pv}. The scalar bulk  field $S$
includes singlet $S_1(x,z)$ and  octet $S_8^a(x,z)$ components,
gathered into the multiplet:
\begin{equation}
S=S^AT^A=S_1T^0+S_8^aT^a
\end{equation}
with $T^0={1}/\sqrt{2n_F}={1}/\sqrt{6}$ and $T^a$ the generators
of $SU(3)_F$  (with  normalization
$\mbox{Tr}\Big(T^AT^B\Big)=\frac{\d^{AB}}{2}$, where
$A=0,a$, and $a=1,\ldots 8$).  $S^A$ is dual to the QCD operator
$\CMcal{O}^A_S(x)=\overline{q}(x)T^A q(x)$,  therefore $m_5^2R^2=-3$ from eq.\eqref{m5}.
The action \eqref{action-new} also involves the fields
$A^a_{L,R}(x,z)$ introduced to gauge the chiral symmetry in the
$5d$ space; they   are dual to the QCD operators $\bar
q_{L,R} \gamma_\mu T^a q_{L,R}$ (defining $\displaystyle q_{L,R}=\frac{1\mp \gamma_5}{2}q$),  with  field strengths:
\begin{equation}
F_{L,R}^{MN}=F_{L,R}^{MNa} T^a=\partial^M A^N_{L,R} - \partial^N A^M_{L,R}-i[A^M_{L,R},A^N_{L,R}]\,\,\, .
\end{equation}
The gauge fields enter in the covariant derivative:
$D^M Y=\partial^M Y-i A_L^M Y +i Y A_R^M$. Writing $A_{L,R}$ in
terms of vector $V$ and axial-vector $A$ fields:
$V^M=\frac{1}{2} (A_L^M+A_R^M)$ and $A^M=\frac{1}{2}
(A_L^M-A_R^M)$,  the action \eqref{action-new} can be written as:
\begin{equation}
S_{eff}=\frac{1}{k^\prime}\int
d^4x dz \,e^{-\phi(z)}\, \sqrt{g}\,\, \mbox{Tr}\Big\{|D Y|^2-m_5^2Y^2-\frac{1}{2g_5^2}\big(F_V^2+F_A^2\big)\Big\}
\label{action1}
\end{equation}
with
\begin{eqnarray}
F_{V}^{MN}&=&\partial^M V^N - \partial^N V^M-i[V^M,V^N]-i[A^M,A^N]\,\,\,,  \non \\
F_{A}^{MN}&=&\partial^M A^N - \partial^N A^M-i[V^M,A^N]-i[A^M,V^N]
\end{eqnarray}
and
$D^M Y=\partial^M Y-i [V^M ,Y] -i \{A^M,Y\}$.

The quadratic part in the field $S^A$ of  this  $5d$ action:
\begin{equation}\label{actionscalar}
S^{(2)}_{S}=\frac{1}{2k^\prime} \int d^4x dz  \,\, e^{-\phi(z)} \sqrt{g} \, \left( g^{MN} \partial_M S^A(x,z)\, \partial_N S^A(x,z) - m_5^2 S^A(x,z) S^A(x,z) \right)
\end{equation}
has been studied  at $T=0$  \cite{Colangelo:2008us},  and can be used to analyze the thermal dependence of the mass of the scalar mesons. The dilaton field is given in \eqref{dilaton} and the AdS-BH metric is used for all temperatures.

Also in this case we define  the bulk-to-boundary propagator  $\tilde S(q,z)$, which satisfies the equation of motion:
\begin{equation}\label{eom2-scalar}
\tilde S''(q,z)- \frac{2c^2z^2f(z)+3+{z^4\over z_h^4}}{z f(z)} \tilde S'(q,z)+{3 \over z^2 f(z)}\tilde S(q,z) +\left( \frac{q_0^2}{f(z)^2}-\frac{\bar{q}^2}{f(z)} \right)  \tilde S(q,z)=0 
\end{equation}
with $q=(q_0,\bar q)$. 

For $\vec q=0$  this equation, written in terms of  the variable  $\displaystyle u={z \over z_h}$, becomes:
\begin{equation}\label{eom3-scalar}
\tilde S''(q_0^2,u)- \frac{2c^2z_h^2 u^2 (1-u^4)+3+u^4}{u (1-u^4)} \tilde S'(q_0^2,u)+{3 \over u^2 (1-u^4)}\tilde S(q_0^2,u)+ \frac{q_0^2 z_h^2} {(1-u^4)^2}   \tilde S(q_0^2,u)=0 
\end{equation}
with the primes denoting again   the derivative with respect to  $u$.
At the horizon $u \to 1$ the independent solutions   of
(\ref{eom3-scalar}) are the same functions as in eq.(\ref{asy1}):
\begin{equation}
\tilde S_\mp(\omega^2,u)=(1-u)^{\mp i \sqrt{\omega^2 z_h^2}/4} 
\end{equation}
so that  we can choose the {\it in falling}  solution as a boundary condition at the horizon, as in \eqref{fallingin}.
For $u \to 0$ eq.\eqref{eom3-scalar}  admits two independent solutions:
\begin{eqnarray}
\tilde S_1(\omega^2,u)&=& u U\left( {2 c^2 -\omega^2 \over 4 c^2}, 0, c^2 z_h^2 u^2\right) \nn \\
\tilde S_2(\omega^2,u)&=& u L\left( -{2 c^2 -\omega^2 \over 4 c^2}, -1, c^2 z_h^2 u^2\right) 
\end{eqnarray}
where $U$ is the Tricomi confluent hypergeometric function  
and $L$ the generalized Laguerre function.
The boundary condition at $u\to0$:  $\tilde S(u)\sim u$ \cite{Colangelo:2008us} allows to write the solution 
\begin{equation}
\tilde S(\omega^2,u)= \tilde S_1(\omega^2,u) + \tilde B(\omega^2) \tilde S_2(\omega^2,u)
\end{equation}
with $\tilde B$, a function of $\omega^2=q_0^2$, numerically  determined as in the case of the scalar glueball.

\begin{figure}[h]
\begin{center}
\includegraphics[width=9cm]{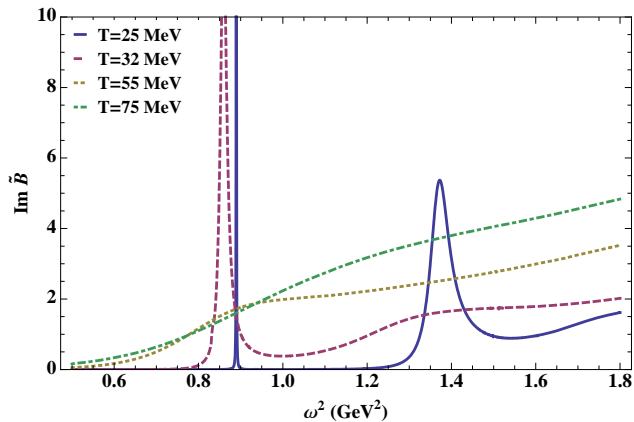}
\caption{Imaginary part of the coefficient $\tilde B(\omega^2)$, proportional to the spectral function Im$\Pi^R_S(\omega^2)$, for  several temperatures in the case of scalar mesons, using  the  AdS-BH metric.}
\label{spfunctionscalar}
\end{center}
\end{figure}

 The retarded two-point Green's function  of the operator $\CMcal{O}^A_S(x)$ can be obtained in terms of  $\tilde S$, and the spectral function ${\rm Im} \Pi^R_S(\omega^2)$ is proportional to the imaginary part of $\tilde B(\omega^2)$.
This spectral function (modulo a numerical overall factor)  is depicted in Fig.\ref{spfunctionscalar} for several values of the temperature. 

\begin{figure}[h]
\subfigure{
\includegraphics[width=6.5cm]{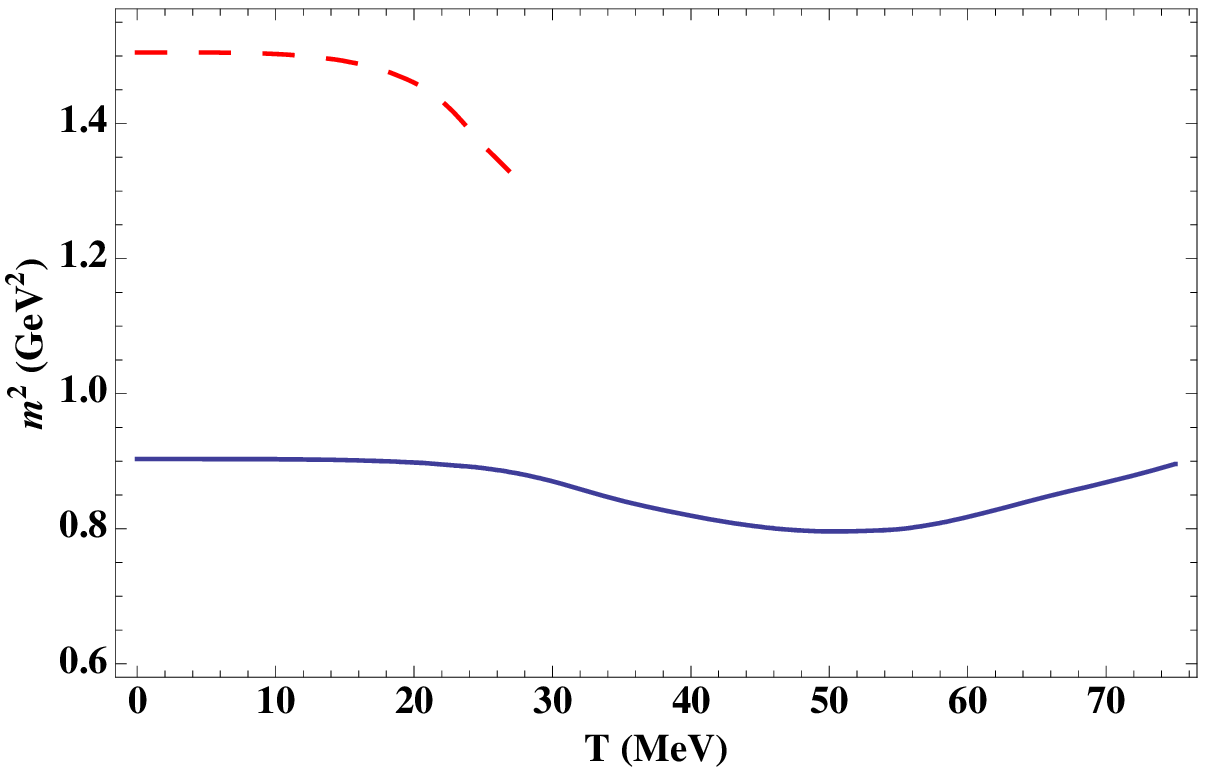}
\label{figmassTscalar}
}
\hspace*{1cm}
\subfigure{
\includegraphics[width=6.5cm]{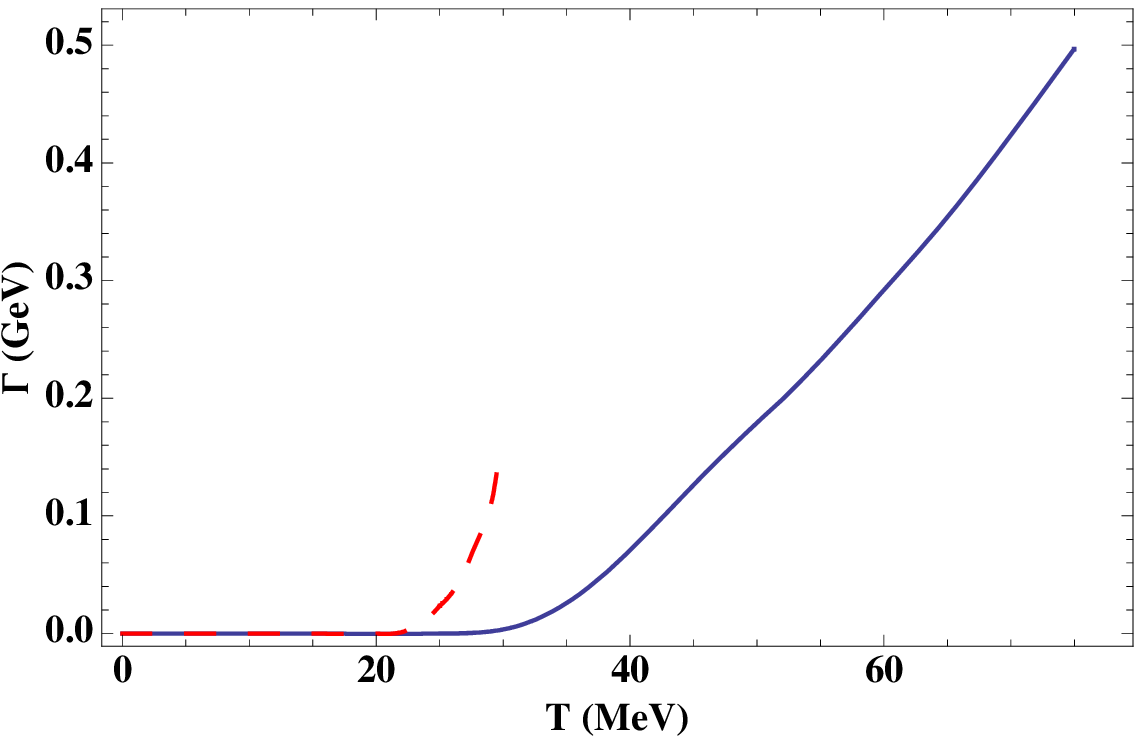}
\label{figwidthTscalar}
}
\caption{Squared mass (left)  and width (right)  of the lightest  scalar meson (continuous lines) and of the first  excited  state (dashed lines)  as a function of the temperature $T$ (MeV) in the SW model with  AdS-BH metric.
Each curve ends  at the temperature where the corresponding peak disappears  from the spectral function.}
\end{figure}

As in the case of  scalar glueballs,  the  spectral function displays  peaks  becoming broader when  the temperature increases. For low $T$, the positions of the peaks correspond to
the spectral condition  $m^2_S=(4 n+6)c^2$ obtained in \cite{Colangelo:2008us},   i.e. $m^2_S=0.913$ GeV$^2$  and $m^2_S=1.521$ GeV$^2$ for the lightest states. Increasing $T$, the masses are shifted towards smaller values, and the
widths  become broader, as shown  in Figs.\ref{figmassTscalar} and \ref{figwidthTscalar} for the first two states. At particular values of the temperature  the peaks disappear from the spectral function.
At odds with the case of  the scalar glueball,   the temperature dependence of the mass of the lowest lying state is milder,  while the dependence on $T$ of the width is visible from $T \simeq 30$ MeV, with an abrupt 
increase with the temperature.  For the first  excitation, the width starts  increasing at $T \simeq 25$ MeV, and for $T \geqslant35$ MeV the peak disappears from the spectral function.

The discussion of these results  follows that presented in the previous Section. The qualitative dependence of masses and widths versus the temperature $T$ agrees with general  expectations, since the particle masses decrease and the widths increase with the temperature. At particular values of $T$ the peaks disappear from the spectral function (melt);   the lightest state survives after the dissolution of the excited states, 
a  behaviour which seems universal in all sectors considered so far. However, also in this case  such phenomena occur at  low temperature   ($T\simeq 40-60$ MeV), unless one invokes once again the presence of a different scale 
($c_{scalar} \neq c_\rho$) to fix the physical  temperatures.  Without such an  assumption,   scalar meson dissociation occurs  in the QCD confined phase,  far from the deconfinement transition.

\section{Models with the Hawking-Page transition}\label{sec:Hawking-Page}

According to the analysis of the minimum of the free energy, in the soft wall model the AdS-BH metric is stable only at high temperatures,  $T \gaq192$ MeV \cite{Herzog:2006ra}. At low temperatures, the stable metric is thermal AdS, and the first order Hawking-Page transition  to the AdS-BH metric is  associated to the deconfinement transition in QCD \cite{Witten:1998zw, Herzog:2006ra}.
Following these hints, the temperature dependence of hadron properties, such as the mass, must be evaluated,  at low temperature, using  thermal AdS metric, while for $T\gaq T_{HP}$  one should use holographic model  based on the 
AdS-BH  metric  described in Sect.\ref{sect:glueball} and \ref{sect:scalar}.

In the case of thermal AdS metric, the equations of motion are the same as at $T=0$;  
from the calculation of the two-point Green's functions and of their spectral functions at   $T\neq0$,  one obtains the  same masses   as at  $T=0$.
Therefore, for temperature   $T$ up to $T_{HP}$, the scalar glueball and scalar meson mass  are given by the spectral formulae derived in \cite{Colangelo:2007pt} and \cite{Colangelo:2008us}.  

\begin{figure}[h]
\subfigure{
\includegraphics[width=6.5cm]{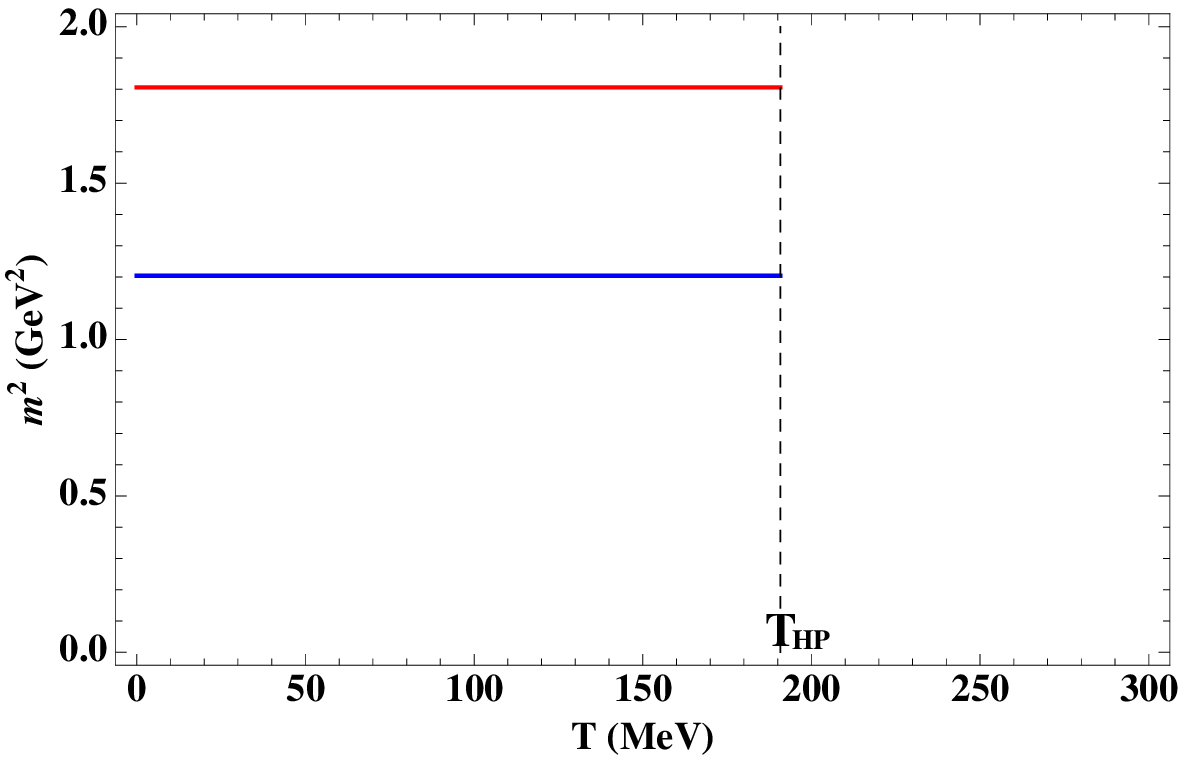}
}
\hspace*{1cm}
\subfigure{
\includegraphics[width=6.5cm]{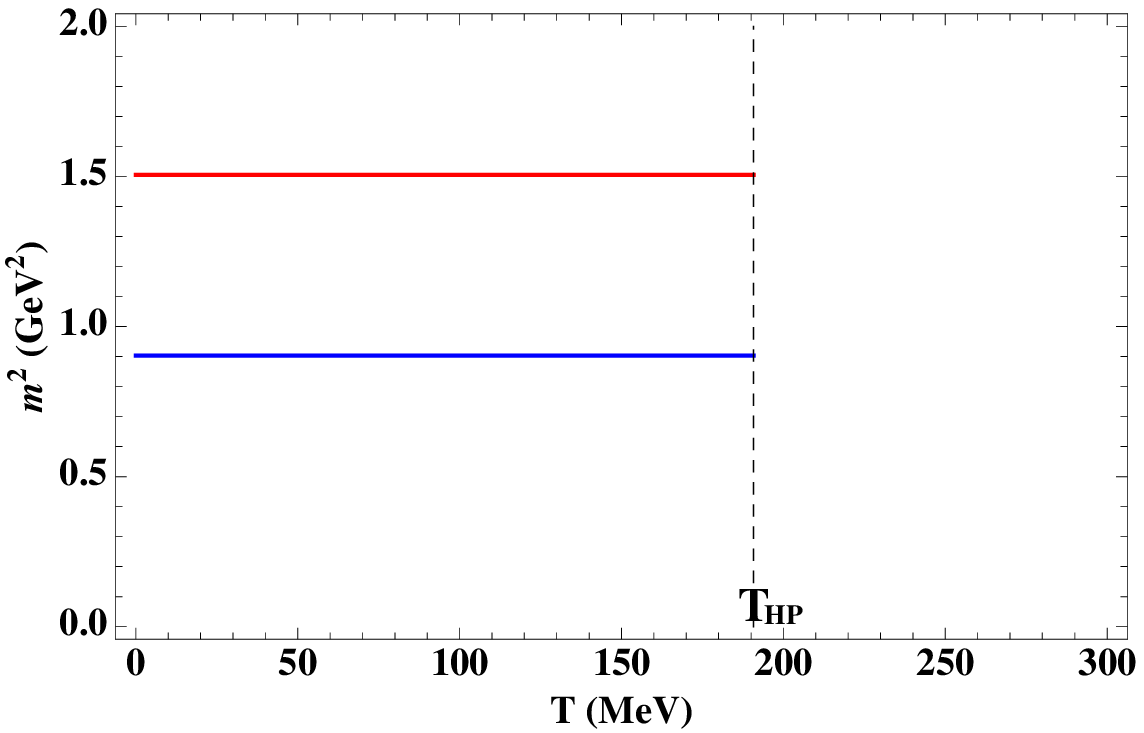}
}
\caption{Squared mass  of the  two lightest scalar glueballs (left)  and  of the  two  lightest scalar mesons (right)   as a function of the temperature $T$ in the soft wall model with HP transition.} \label{HP_masses}
\end{figure}

On the other hand, for $T\geqslant T_{HP}$  the results in Sect.\ref{sect:glueball} and \ref{sect:scalar}  show
 that in both  the scalar  glueball and scalar meson sectors  no peaks appear in the spectral functions:   dissociation has already occurred at these temperatures. At $T=T_{HP}$, when  the black hole appears  in the metric, the masses   jump from $m^2 \neq 0$ to $m^2=0$:   dissociation occurs together with deconfinement, as it could be expected in a discontinuous transition. This is shown in Fig. \ref{HP_masses}. 
 As observed in \cite{Herzog:2006ra}, the temperature independence of the mass spectrum below $T_c$ is consistent with large $N_c$ expectations, and is supported by chiral perturbation theory analyses \cite{ioffe}.
 
The same conclusion holds for the vector mesons considered in \cite{Fujita:2009wc}:   the temperature dependence of the vector meson mass and the broadening of the width in a metastable phase,
such as AdS-BH at $T\leqslant T_{HP}$, are different than in the  stable phase,  and  a model aimed at describing  QCD should  take the difference into account. However,  in modifications of the soft wall model  as proposed, e.g.,  in \cite{Kwee:2007nq}, the possibility of the persistence of some hadron resonances   above $T_c$  is not excluded, and needs to  be investigated by  a dedicated study.

\section{Conclusions}
Without considering the existence of the HP critical point,  the SW model is able to reproduce  only qualitatively  some commonly expected features of finite temperature QCD,  like  in-medium mass  shifts and  width broadening,  but
at  temperatures  different from those found  by  lattice QCD simulations. Using  the  AdS-BH for all values of $T$  would imply that scalar glueballs and scalar mesons disappear
from the spectral functions (melt) at  temperatures  of about $40-60$~MeV.  On the other hand,  in a  holographic description  based on SW with thermal AdS geometry below the  
Hawking-Page transition temperature, and  AdS-BH geometry above this temperature,   the hadronic states are found to persist in the confined phase  and melt at  deconfinement.  
This suggests  that a more refined dual model of finite temperature QCD could be found modifying the background dilaton, so that the qualitative behaviour is preserved while  the temperature scale is enlarged.

\appendix*
\section{Scalar glueball at  $T\neq 0$ in  the hard wall model }
It is interesting to consider  scalar glueballs at finite $T$ in the hard wall holographic model of QCD. In this model  an AdS slice is used, up to a maximum value of $z$, $z_m$, and there is no dilaton-like background field.
Therefore,    it is sufficient to put $c=0$ in the equations obtained in Sect.\ref{sect:glueball}, imposing suitable boundary conditions.
Using the AdS-BH metric  for all values of $T$,  as done in an analysis of the static potential in QCD \cite{andreev1},
two cases are possible.  In the first one   the black hole horizon  is located beyond the IR cutoff $z_m$: $z_m<z_h$; in the second  one   $z_h<z_m$.  In the first case, regardless of position of the  horizon,  the eigenfunctions can be obtained   solving  the equation of motion with boundary conditions:
$$X(m^2,0)=0 \qquad\qquad X'(m^2,z_m)=0$$
hence determining the  mass squared of the scalar glueballs.

\begin{figure}[h]
\begin{center}
\includegraphics[width=7cm]{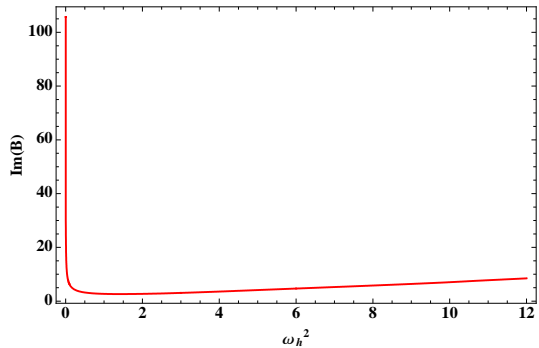}
\caption{Imaginary part of the coefficient $B(\omega^2_h)$,  proportional to the spectral function  $Im \Pi^R_G(\omega^2_h)$, in the hard wall model with AdS-BH metric ,  for $z_h<z_m$. A peak  is found  only at $\omega_h^2=0$. }
\label{figrhoHW}
\end{center}
\end{figure}

\begin{figure}[h]
\begin{center}
\includegraphics[width=7cm]{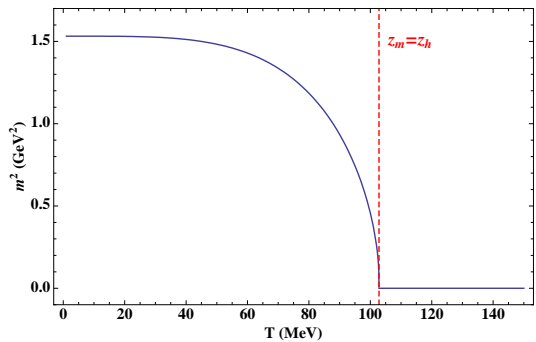}
\caption{Squared mass of the lightest scalar glueball as a function of the  temperature $T$ in the hard wall model with AdS-BH.}
\label{figM2vsTHW}
\end{center}
\end{figure}

When $z_m>z_h$, the horizon position $z_h$ becomes the only mass scale,  and the equation of motion reads:
\begin{equation}\label{eom4HW}
\tilde K''(m_h^2,u)- \frac{3+u^4}{u (1-u^4)} \tilde K'(m_h^2,u)+ \frac{m_h^2}{(1-u^4)^2}  \tilde K(m_h^2,u)=0    \quad\quad 0<u<1\,
\end{equation}
where $m_h=m z_h$.
The masses can be  obtained imposing  that the solution of \eqref{eom4HW} is  the {\it in falling  } solution into the  black hole at  $u\rightarrow1$ and considering spectral function of the  retarded Green's function.  Once having obtained $m_h^2$, 
the  glueball mass is given  by $m^2=m_h^2 \pi^2 T^2$,  hence  non vanishing values of $m$  linearly increase  with the temperature $T$.
However,  the spectral function depicted in  Fig.~\ref{figrhoHW}  has only one peak at $m_h^2=0$;  therefore,  when  $z_h<z_m$ (or  $T \geqslant 1/(\pi z_h)$)  the only value found for the scalar glueball mass  is  $m^2=0$.
The resulting plot  of the squared masses  at various temperatures  $T$ is shown in Fig.\ref{figM2vsTHW}:   dissociation occurs when $z_m=z_h$. 
This corresponds to $T\simeq 103$ MeV, using the value of $z_m$ fixed from the mass of the $\rho$ meson  in the HW  model: $z_m=\frac{1}{323}$ MeV$^{-1}$   \cite{son1}.  Analogous results  hold  for  mesons  in HW 
at $T \neq 0$ \cite{Ghoroku:2005kg}.

Imposing the presence of the Hawking-Page transition, which  occurs at $T_{HP}=2^{1/4}/(\pi z_m)\sim 122$ MeV,   i.e. for $z_h<z_m$,   the masses do not vary up to the critical temperature $T_{HP}$,  at which they jump to  $m^2=0$.
After completing this work, we noticed the preprint \cite{braga1}, where an
analysis of glueballs at finite temperature similar to the one presented
here has been carried out.

\begin{acknowledgments}
 \noindent
One of us, PC,  thanks  T.N. Pham  and C. Roiesnel  for discussions and for warm hospitality at the Centre de Physique Th\'eorique,  \'Ecole Polytechnique, CNRS,  Palaiseau, France.   We are grateful to F. De Fazio,  M. Pellicoro and E. Scrimieri for discussions. This work  was supported in part by the EU contract No. MRTN-CT-2006-035482, "FLAVIAnet".
 \end{acknowledgments}


\newpage
\begin{thebibliography}{99}

\bibitem{rassegna}
For recent reviews see:
  H.~Satz,
  J.\ Phys.\ G {\bf 32}, R25 (2006);
  N.~Armesto {\it et al.},
  J.\ Phys.\ G {\bf 35}, 054001 (2008).

\bibitem{Ding:2009se}
  H.~T.~Ding, O.~Kaczmarek, F.~Karsch and H.~Satz,
  PoS  {\bf CONFINEMENT8}, 108 (2008) and references therein.

\bibitem{rassegna1}
Examples of calculation of hadronic properties at finite temperature are discussed in the  references:\\
for lattice QCD: 
  M.~Asakawa and T.~Hatsuda,
  Phys.\ Rev.\ Lett.\  {\bf 92}, 012001 (2004); \\
for QCD sum rules:   K.~Morita and S.~H.~Lee,
  arXiv:0908.2856 [hep-ph];
  C.~A.~Dominguez, M.~Loewe, J.~C.~Rojas and Y.~Zhang,
  arXiv:0908.2709 [hep-ph]; \\
 for potential models: A.~Mocsy,
  Eur.\ Phys.\ J.\  C {\bf 61}, 705 (2009),  and in reference therein.

\bibitem{Maldacena:1997re}
  J.~M.~Maldacena,
  Adv.\ Theor.\ Math.\ Phys.\  {\bf 2}, 231 (1998)
  [Int.\ J.\ Theor.\ Phys.\  {\bf 38}, 1113 (1999)].

\bibitem{Witten:1998qj}
  E.~Witten,
  Adv.\ Theor.\ Math.\ Phys.\  {\bf 2}, 253 (1998).

\bibitem{Gubser:1998}
  S.~S.~Gubser, I.~R.~Klebanov and A.~M.~Polyakov,
  Phys.\ Lett.\  B {\bf 428},  105 (1998).
  
\bibitem{Witten:1998zw}
  E.~Witten,
  Adv.\ Theor.\ Math.\ Phys.\  {\bf 2}, 505 (1998).
  
  \bibitem{HPage}
  S. W.~Hawking and D. N. Page,
  Commun.\ Math.\ Phys.\  {\bf 87}, 577 (1983).
  
  \bibitem{teper}
   B.~Lucini, M.~Teper and U.~Wenger,
  JHEP {\bf 0502}, 033 (2005).

\bibitem{erdmenger}
  R.~C.~Myers, A.~O.~Starinets and R.~M.~Thomson,
  JHEP {\bf 0711}, 091 (2007);
  J.~Erdmenger, M.~Kaminski and F.~Rust,
  PoS  {\bf CONFINEMENT8}, 131 (2008).

\bibitem{polchinsky}
  J.~Polchinski and M.~J.~Strassler,
  Phys.\ Rev.\ Lett.\  {\bf 88},  031601 (2002).
  
  \bibitem{son1}
    J.~Erlich, E.~Katz, D.~T.~Son and M.~A.~Stephanov,
  Phys.\ Rev.\ Lett.\  {\bf 95},  261602 (2005).

\bibitem{pomarol1}
   L.~Da Rold and A.~Pomarol,
  Nucl.\ Phys.\  B {\bf 721}, 79 (2005).

\bibitem{teramond1}
     S.~J.~Brodsky and G.~F.~de Teramond,
  Phys.\ Rev.\ Lett.\  {\bf 94},  201601 (2005),
  Phys.\ Rev.\ Lett.\  {\bf 96},  201601 (2006).
  
\bibitem{Grigoryan:2007vg}
  H.~R.~Grigoryan and A.~V.~Radyushkin,
  Phys.\ Lett.\  B {\bf 650}, 421 (2007).

\bibitem{Karch:2006pv}
  A.~Karch, E.~Katz, D.~T.~Son and M.~A.~Stephanov,
  Phys.\ Rev.\  D {\bf 74}, 015005 (2006).

\bibitem{Andreev:2006vy}
  O.~Andreev,
  Phys.\ Rev.\  D {\bf 73}, 107901 (2006).
 
\bibitem{Herzog:2006ra}
  C.~P.~Herzog,
  Phys.\ Rev.\ Lett.\  {\bf 98}, 091601 (2007).

\bibitem{braga}
  C.~A.~Ballon Bayona, H.~Boschi-Filho, N.~R.~F.~Braga and L.~A.~Pando Zayas,
  Phys.\ Rev.\  D {\bf 77}, 046002 (2008).
  
\bibitem{Gursoy:2009jd}
  U.~Gursoy, E.~Kiritsis, L.~Mazzanti and F.~Nitti,
  Phys.\ Rev.\ Lett.\  {\bf 101}, 181601 (2008);
  JHEP {\bf 0905}, 033 (2009);
  Nucl.\ Phys.\  B {\bf 820}, 148 (2009).

\bibitem{Fujita:2009wc}
  M.~Fujita, K.~Fukushima, T.~Misumi and M.~Murata,
    Phys.\ Rev.\  D {\bf 80}, 035001 (2009).
  
\bibitem{Colangelo:2007pt}
  P.~Colangelo, F.~De Fazio, F.~Jugeau and S.~Nicotri,
  Phys.\ Lett.\  B {\bf 652}, 73 (2007);
%
  Int. \ J. \ Mod. \ Phys. \ A {\bf 24},  4177 (2009).
 
\bibitem{Colangelo:2008us}
  P.~Colangelo, F.~De Fazio, F.~Giannuzzi, F.~Jugeau and S.~Nicotri,
  Phys.\ Rev.\  D {\bf 78}, 055009 (2008).
 
\bibitem{Teaney:2006nc}
  D.~Teaney,
  Phys.\ Rev.\  D {\bf 74}, 045025 (2006).

  \bibitem{starinets}
 D.~T.~Son and A.~O.~Starinets,
  JHEP {\bf 0209}, 042 (2002);
  G.~Policastro, D.~T.~Son and A.~O.~Starinets,
  JHEP {\bf 0209}, 043 (2002).
   
  \bibitem{lattice}
  N.~Ishii, H.~Suganuma and H.~Matsufuru,
  Phys.\ Rev.\  D {\bf 66}, 014507 (2002);
  Phys.\ Rev.\  D {\bf 66}, 094506 (2002);
   X.~F.~Meng, G.~Li, Y.~Chen, C.~Liu, Y.~B.~Liu, J.~P.~Ma and J.~B.~Zhang,
  arXiv:0903.1991 [hep-lat].

 \bibitem{ioffe}
  V.~L.~Eletsky and B.~L.~Ioffe,
  Phys.\ Rev.\  D {\bf 51}, 2371 (1995).
  
\bibitem{Kwee:2007nq}
  H.~J.~Kwee and R.~F.~Lebed,
  Phys.\ Rev.\  D {\bf 77}, 115007 (2008);
  T.~Gherghetta, J.~I.~Kapusta and T.~M.~Kelley,
  Phys.\ Rev.\  D {\bf 79}, 076003 (2009).
  
\bibitem{andreev1}
  O.~Andreev and V.~I.~Zakharov,
  Phys.\ Lett.\  B {\bf 645}, 437 (2007).
  
\bibitem{Ghoroku:2005kg}
  K.~Ghoroku and M.~Yahiro,
  Phys.\ Rev.\  D {\bf 73}, 125010 (2006).

\bibitem{braga1}
  A.~S.~Miranda, C.~A.~B.~Bayona, H.~Boschi-Filho and N.~R.~F.~Braga,
  arXiv:0909.1790 [hep-th].
\end{thebibliography}
\end{document}